\documentclass[twocolumn,showpacs,superscriptaddress]{revtex4}
\usepackage{graphicx}
\usepackage{amssymb}
\usepackage{amsmath}

\newcommand{\bk}{{\bf k}}

\newcommand{\br}{{\bf r}}

\newcommand{\sgn}{{\mathop{\rm{sgn}}\nolimits\,}}

\newcommand{\kB}{k_{\mathrm{B}}}
\newcommand{\kF}{k_{\mathrm{F}}}

\begin{document}

\title{Statistical correlations in an ideal gas of particles obeying fractional
exclusion statistics}

\author{F. M. D. Pellegrino}
\affiliation{Scuola Superiore di Catania, Via S. Nullo, 5/i, I-95123 Catania,
Italy}
\author{G. G. N. Angilella}
\email[Corresponding author. E-mail: ]{giuseppe.angilella@ct.infn.it}
\affiliation{Dipartimento di Fisica e Astronomia, Universit\`a di Catania,\\ and
CNISM, UdR Catania, and INFN, Sez. Catania,\\ Via S. Sofia, 64, I-95123 Catania,
Italy}
\author{N. H. March}
\affiliation{Department of Physics, University of Antwerp,\\
Groenenborgerlaan 171, B-2020 Antwerp, Belgium}
\affiliation{Oxford University, Oxford, England} 
\author{R. Pucci}
\affiliation{Dipartimento di Fisica e Astronomia, Universit\`a di Catania,\\ and
CNISM, UdR Catania, and INFN, Sez. Catania,\\ Via S. Sofia, 64, I-95123 Catania,
Italy}

\date{\today}

\begin{abstract}
\medskip
After a brief discussion of the concepts of fractional exchange and fractional
exclusion statistics, we report partly analytical and partly numerical results
on thermodynamic properties of assemblies of particles obeying fractional
exclusion statistics. The effect of dimensionality is one focal point, the ratio
$\mu/\kB T$ of chemical potential to thermal energy being obtained numerically
as a function of a scaled particle density. Pair correlation functions are also
presented as a function of the statistical parameter, with `Friedel'
oscillations developing close to the fermion limit, for sufficiently large
density.\\
\end{abstract} 

\maketitle

\section{Introduction}

Current interest in fractional statistics
\cite{Leinaas:77,Wilczek:82,Wilczek:82a} is motivated by its possible relevance
for fractional quantum Hall effect \cite{Laughlin:83,Laughlin:83a} and
high-temperature superconductivity \cite{Laughlin:88}.  In this context, an
early application was made by Lea \emph{et al.} \cite{Lea:91,Lea:92} to discuss
the de~Haas--van~Alphen oscillatory orbital magnetism of a two-dimensional
electron gas (2DEG). Quite recently, it has been proposed that noise experiments
in quantum Hall fluids might reveal the existence of elementary excitations
obeying fractional statistics \cite{Kim:05,Camino:05}. 

Fractional \emph{exchange} statistics \cite{Leinaas:77,Wilczek:82,Wilczek:82a}
arises when the many-body wave function of a system of indistinguishable
particles (dubbed anyons) is allowed to acquire an arbitrary phase
$e^{i\alpha\pi}$ upon an adiabatic exchange process of two particles. Here,
$\alpha$ is the so-called statistical parameter, interpolating between
$\alpha=0$ (bosons) and $\alpha=1$ (fermions). Such an exchange produces a
nontrivial phase only if the configuration space of the collection of particles
under study possesses a multiply connected topological structure. Therefore,
fractional exchange statistics is usually restricted to two spatial dimensions,
$d=2$ (see Ref.~\cite{Forte:92} for a review). However, fractional exchange
statistics can be formalized, to some extent, also in $d=1$
\cite{Ha:96,Calabrese:07}.

A different concept of fractional statistics, namely fractional \emph{exclusion}
statistics, is based on the structure of the Hilbert space, rather than the
configuration space, of the particle assembly, and is therefore not restricted
to $d\leq2$. Fractional exclusion statistics has been introduced by Haldane
\cite{Haldane:91}, who considered the ratio
\begin{equation}
g=-\Delta D/\Delta N,
\label{eq:Haldane}
\end{equation}
where $\Delta D$ denotes the change in size of the subset of available states in
the Hilbert space corresponding to a change $\Delta N$ of the number of
particles (\emph{i.e.} elementary excitations). Clearly, one has again $g=0$ for
bosons and $g=1$ for fermions, the latter being a consequence of Pauli exclusion
principle. At variance with anyons, particles obeying fractional exclusion
statistics are usually dubbed $g$-ons or exclusons. The distribution function
for fractional exclusion statistics has been derived by Wu \cite{Wu:94}. A
limiting form of the same distribution function had been derived by March
\emph{et al.} \cite{March:93c} within an approximate chemical collision model
(see also Refs.~\cite{March:93b,March:97b}). Following Wu's distribution function
\cite{Wu:94}, the thermodynamics of an ideal gas with fractional exclusion
statistics has been studied in some detail in arbitrary dimensions
\cite{Joyce:96,Iguchi:97}.

The relation between fractional exchange and exclusion statistics is elusive. In
particular, it has been emphasized \cite{Nayak:94} that in order to derive a
consistent statistical mechanics for $g$-ons, Haldane's generalized exclusion
principle, Eq.~(\ref{eq:Haldane}), must hold locally in phase space, which
indeed applies rigorously to either bosons or fermions. In other words, $\Delta
D$ in Eq.~(\ref{eq:Haldane}) should be related to states of close energy, which
is brought about by an effective interaction which is local in momentum space.
From this, it has been concluded that anyons are not ideal $g$-ons, but
interacting $g$-ons \cite{Nayak:94}. Haldane \cite{Haldane:91} provides explicit
examples of systems for which fractional exclusion, albeit not exchange,
statistics might be applicable. An earlier attempt to relate the parameter $g$
of fractional exclusion statistics and the statistical parameter $\alpha$ of
fractional exchange statistics has been made in Ref.~\cite{Murthy:94}, on the
basis of the virial expansion of the $g$-ons partition function. More recently,
an analytic, monotonic relation $g=g(\alpha)$ has been derived in
Ref.~\cite{Speliotopoulos:97} in the case $g=1/m$ (for integer $m$), which is
relevant for the fractional quantum Hall effect (see also
Ref.~\cite{Canright:94} for a review).

In an electron gas, statistical correlations between non-interacting fermions
arise as a consequence of particle indistinguishability. In the case of
fermions, the antisymmetry of the wave function implies that two such particles
can never be found simultaneously at the same point in space. As a result, each
electron is surrounded by an `exchange hole', a region in which the density of
same-spin electrons is smaller than average and in which, therefore, the
positive background charge is not exactly cancelled. It is the interaction of
each electron with the positive charge of the exchange hole that gives rise to
the exchange energy. The situation is quite the opposite for bosons, where
statistical correlations are attractive and may be thought of being ultimately
responsible of Bose-Einstein condensation. In the case of intermediate or
fractional statistics, the question naturally arises of how many particles does
a particle effectively `see' around itself. In other words, one is led to
consider the effect of Haldane's generalized exclusion principle on the exchange
hole which a $g$-on `digs' around itself.

In this work, we attempt to answer such a question, by evaluating a suitably defined
generalization of the pair correlation function between $g$-ons (\emph{i.e.}
particles obeying fractional \emph{exclusion} statistics), in arbitrary
dimensions $d$ ($d=1,2,3$), as a function of temperature $T$ and particle
density. Earlier results along this direction include the seminal work of
Sutherland \cite{Sutherland:71}, for a system of fermions or bosons in $d=1$
interacting through a singular potential $\sim 1/r^2$, a problem which can be
mapped into anyons \cite{Forte:92}. The evaluation of a pairwise correlation
function for anyons (\emph{i.e.} particles obeying fractional \emph{exchange}
statistics) in $d=2$ at $T=0$ has been also recently analyzed in
Ref.~\cite{Gutierrez:04}, in connection with an intensity interferometry
gedanken experiment, which may prove useful in the context of quantum computing.
The question is also intimately related to that of finding Uhlenbeck's
statistical interparticle potential \cite{Uhlenbeck:32}, which has been studied
for the two-anyons system in $d=2$ \cite{Huang:95}.

The paper is organized as follows. In Sec.~\ref{sec:paircorr} we derive the pair
correlation function for a homogeneous assembly of $g$-ons as a function of
interparticle separation $r$. In Sec.~\ref{ssec:chempot} we study numerically
the dependence of the chemical potential on density at a given temperature, for
arbitrary dimensions $d=1,2,3$, thereby reproducing Wu's analytical result at
$d=2$. In Sec.~\ref{ssec:paircorr} we study numerically the pair correlation
function, and in particular focus on the possible occurrence of `Friedel'
oscillations, depending on the statistical parameter and on the scaled density.
We eventually summarize in Sec.~\ref{sec:summary}.

\section{Pair correlation function for non-interacting $g$-ons}
\label{sec:paircorr}

Statistical correlations between indistinguishable, hard-core particles can be
described by means of the pair correlation function $g(\br)$, defined as the
normalized probability of simultaneously finding a particle at position $\br$
and a particle at position $\br=0$. (Here and below, we are implicitly
neglecting spin.) In the formalism of second quantization, the pair correlation
function can be written as 
\begin{equation}
g(\br) = \frac{\langle \Psi^\dag (\br) \Psi^\dag (0) \Psi (0)
\Psi(\br)\rangle}{n(\br) n(0)} .
\label{eq:gr}
\end{equation}
Here, $\Psi^\dag (\br)$ [$\Psi (\br)$] is a creation [annihilation] quantum
field operator at position $\br$, $n(\br) = \langle \Psi^\dag (\br) \Psi
(\br)\rangle$ is the single-particle probability density at position $\br$, and
$\langle\ldots\rangle$ denotes a quantum statistical average associated with the
equilibrium distribution of the particle assembly under study (see below). In
the case of a homogeneous system, one obviously has $n(\br)=n(0)$ and
$g(\br)\equiv g(r)$.

In the case of either fermions or bosons, the four-point average appearing in
the right-hand side of Eq.~(\ref{eq:gr}) can be reduced by means of Wick's
theorem as
\begin{eqnarray}
\langle \Psi^\dag (\br) \Psi^\dag (0) \Psi (0) \Psi(\br)\rangle &=&
\langle \Psi^\dag (\br) \Psi(\br) \rangle 
\langle \Psi^\dag (0) \Psi(0) \rangle \nonumber \\
&&\pm
\langle \Psi^\dag (\br)\Psi(0) \rangle\langle \Psi^\dag (0)\Psi(\br) \rangle,
\nonumber \\
\label{eq:Wick}
\end{eqnarray}
the plus (minus) sign corresponding to bosons (fermions). In the case of
particles obeying exclusion statistics, the appropriate generalization of
Wick's theorem depends on the behavior of excluson field operators under
exchange. As emphasized in the Introduction, the relation between fractional
exclusion and exchange statistics is not completely settled. Within fractional
exchange statistics in reduced dimensionality ($d\leq 2$), one obviously has the
following graded commutation relations,
\begin{subequations}
\begin{eqnarray}
[\Psi(\br),\Psi(\br^\prime)]_\alpha &\equiv& \Psi(\br)\Psi(\br^\prime) -
e^{i\alpha\pi s(\br, \br^\prime )} \Psi(\br^\prime)\Psi(\br)
\nonumber \\
 &=& 0,\\
{}[\Psi^\dag(\br),\Psi^\dag(\br^\prime)]_\alpha  &=& 0,\\
{}[\Psi(\br),\Psi^\dag(\br^\prime)]_\alpha  &=&
\delta(\br-\br^\prime),
\end{eqnarray}
\label{eq:comm}
\end{subequations}
with $s(\br, \br^\prime ) =\pm 1$, depending on whether the exchange is being
performed along a (counter)clockwise path in $d=2$ \cite{Forte:92}, or $s(\br,
\br^\prime ) = \sgn (r-r^\prime)$ in $d=1$ \cite{Ha:96}. Eqs.~(\ref{eq:comm})
naturally reduce to the familiar commutation and anticommutation relations for
bosons ($\alpha=0$) and fermions ($\alpha=1$), respectively. In $d=3$, the sign
$s(\br, \br^\prime )$ is ill-defined, and in fact fractional \emph{exchange}
statistics does not apply to dimensions $d\geq 3$. In order to have generalized
commutation relations valid for exclusons in arbitrary dimensions, we therefore
assume real field operators, assign a statistical exchange parameter $\alpha$ to
exclusons ($g$-ons), and postulate that standard quantum Bose or Fermi
statistics are weakly violated as in
\begin{subequations}
\begin{eqnarray}
[\Psi(\br),\Psi(\br^\prime)]_\alpha &\equiv& \Psi(\br)\Psi(\br^\prime) -
\cos(\alpha\pi) \Psi(\br^\prime)\Psi(\br)
\nonumber \\
 &=& 0,\\
{}[\Psi^\dag(\br),\Psi^\dag(\br^\prime)]_\alpha  &=& 0,\\
{}[\Psi(\br),\Psi^\dag(\br^\prime)]_\alpha  &=&
\delta(\br-\br^\prime),
\end{eqnarray}
\label{eq:comm-cos}
\end{subequations}
which again reduce to the familiar relations for bosons and fermions, in the
appropriate limits. Following Ref.~\cite{Greenberg:91} for the generalization of
Wick's theorem corresponding to Eqs.~(\ref{eq:comm-cos}), Eq.~(\ref{eq:Wick})
becomes
\begin{eqnarray}
\langle \Psi^\dag (\br) \Psi^\dag (0) \Psi (0) \Psi(\br)\rangle &=&
\langle \Psi^\dag (\br) \Psi(\br) \rangle 
\langle \Psi^\dag (0) \Psi(0) \rangle \nonumber \\
&&\!\!\!\!\!\!\!\!\!\!\!\!\!\!\!\!\!\!\!\!\!\!\!\!\!\!\!\!\!\!\!\!\!\!\!\!\!\!\!\!
+\cos(\alpha\pi)
\langle \Psi^\dag (\br)\Psi(0) \rangle\langle \Psi^\dag (0)\Psi(\br) \rangle.
\nonumber \\
\end{eqnarray}
Correspondingly, the pair correlation function for a homogeneous assembly of
$g$-ons reads
\begin{equation}
g (r ) = 1+\cos(\alpha\pi) 
\frac{|\langle \Psi^\dag (r )  \Psi (0 )
\rangle|^2}{n^2 (0)} ,
\label{eq:galpha}
\end{equation}
which correctly reduces to the fermion limit ($\alpha=1$), with $g(r)\leq 1$,
manifestly \cite{Giuliani:05}.

For translationally invariant systems, one may use a plane wave expansion to
find
\begin{equation}
g (r) = 1+\cos(\alpha\pi) \left| \frac{\tilde{n} (r)}{\tilde{n} (0)}
\right|^2 ,
\label{eq:gnFourier}
\end{equation}
where
\begin{equation}
\tilde{n}(r) = \int \frac{d^d \bk}{(2\pi)^d} e^{-i\bk\cdot\br} n(\bk)
\label{eq:gnFourier1}
\end{equation}
is the Fourier transform in $d$ dimensions of the single-particle distribution
function $n(\bk)$ for $g$-ons in equilibrium at temperature $T$.

Following Wu \cite{Wu:94}, the average occupation number $n_i$ of exclusons in
an energy state $\epsilon_i$ at equilibrium is
\begin{equation}
n_i = \frac{1}{w(\zeta ) + \alpha} ,
\label{eq:Wun}
\end{equation}
where $\zeta = \exp [(\epsilon_i -\mu)/\kB T]$, $\mu$ is the chemical potential,
and the function $w(\zeta)$ satisfies the functional equation
\begin{equation}
w^\alpha (1+w)^{1-\alpha} = \zeta .
\end{equation}
Here and below, we shall use of the standard notation $g\equiv\alpha$, even for
exclusons. Eq.~(\ref{eq:Wun}) establishes the thermodynamics of the excluson
assembly, once the energy level distribution $\epsilon_i$ is given, which is
determined by the free-particle dynamics. In order to make contact with the
ordinary Bose and Fermi limits, we shall then assume that $\epsilon_i = \hbar^2
k_i^2 /(2m)$, and then treat $\bk_i \mapsto \bk$ as a continuous variable, as is
understood in Eq.~(\ref{eq:gnFourier}) with $n_i \mapsto n(\bk)$.

An approximate form of the distribution function for exclusons had been earlier
derived by March \emph{et al.} \cite{March:93c,March:93b,March:97b} from
collision theory, by using the detailed balance hypothesis. Their result was
that $n_i^{-1} = \zeta + a$, with $a$ interpolating between Bose ($a=-1$) and
Fermi ($a=+1$) statistics. Comparison with Wu's result, Eq.~(\ref{eq:Wun}), then
yields \cite{March:93c,March:97b}
\begin{equation}
a(\alpha,w) = w + \alpha - w^\alpha (1+w)^{1-\alpha} \approx 2\alpha -1,
\label{eq:Marchexpansion}
\end{equation}
the latter approximation, independent of $w$, holding in the limit $w\gg 1$.
Eq.~(\ref{eq:Marchexpansion}) already correctly interpolates between the Bose
($\alpha=0$, $a=-1$) and Fermi ($\alpha=1$, $a=1$) limits. Of course, as $w$
becomes small, the dependence of $a$ on $w$ is significant, though $2\alpha-1$
is not a bad approximation even at $w=1$ (giving for $\alpha=\frac{1}{2}$, $a=0$
instead of $0.09$, and for $\alpha=\frac{1}{4}$, $a=-\frac{1}{2}$ rather than
$-0.44$).

\subsection{Chemical potential}
\label{ssec:chempot}

Within these assumptions, the general properties and some exact results of the
thermodynamics of an assembly of noninteracting exclusons have been derived to
some extent \cite{Wu:94,Joyce:96,Iguchi:97}. Some properties of Wu's function
$w(\zeta)$ in the complex $\zeta$-plane and their relevance for Cooper pairing
of exclusons have been discussed in Ref.~\cite{Angilella:06a}. In particular, by
evaluating the particle density
\begin{equation}
\frac{N}{V} = \int \frac{d^d \bk}{(2\pi)^d} n(\bk)
\label{eq:mud}
\end{equation}
in dimensions $d=2$, Wu \cite{Wu:94} was able to derive the relation between chemical
potential $\mu$, temperature $T$ and particle density $N/V$ explicitly as
\begin{equation}
\frac{\mu}{\kB T} = \alpha \frac{\lambda^2 N}{V} + \log \left[ 1 - \exp \left(
-\frac{\lambda^2 N}{V} \right) \right] ,
\end{equation}
where $\lambda = \hbar \sqrt{2\pi/(m\kB T)}$ is the thermal wavelength.
Performing the integration in Eq.~(\ref{eq:mud}) for a homogenous system in
arbitrary dimensions $d$ (see also Ref.~\cite{Iguchi:97}), such a relation can
be generalized implicitly as
\begin{equation}
\Gamma \left( \frac{d}{2} \right) \frac{\lambda^d N}{V} = \int_{w_0}^\infty dw
\, \frac{1}{w(1+w)}  \bar{\epsilon}^{\frac{d-2}{2}} , 
\label{eq:mud1}
\end{equation}
where $\Gamma(x)$ is Euler's function, $w_0 \equiv w_0 (\mu/\kB T)$ is a
generalized inverse fugacity, implicitly defined by Wu's functional equation for
$\epsilon=0$, \emph{i.e.}
\begin{equation}
w_0^\alpha (1+w_0 )^{1-\alpha} = e^{-\mu/\kB T} ,
\label{eq:w0}
\end{equation}
and
\begin{equation}
\bar{\epsilon} = \log \left[ w^\alpha (1+w)^{1-\alpha} e^{\mu/\kB T}
\right] \equiv \log \frac{w^\alpha (1+w)^{1-\alpha}}{w_0^\alpha (1+w_0
)^{1-\alpha}} .
\label{eq:walpha}
\end{equation}

\begin{figure}[t]
\centering
\includegraphics[height=0.9\columnwidth,angle=-90]{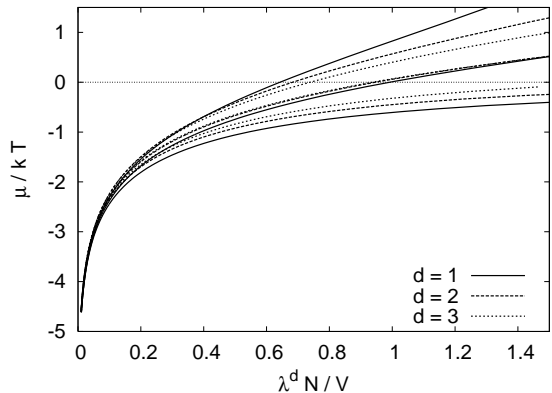}
\caption{Chemical potential \emph{vs} scaled density, Eq.~(\ref{eq:mud1}), for
$\alpha=0,\, 0.5,\, 1$ (bottom to top) and $d=1,2,3$.}
\label{fig:mu}
\end{figure}

Eliminating $w_0$ between Eqs.~(\ref{eq:mud1}) and (\ref{eq:w0}), we obtain the
relation between the ratio $\mu/\kB T$ of chemical potential to thermal energy
and the scaled density $\lambda^d N/V$, parametrically. This is plotted in
Fig.~\ref{fig:mu} for several values of the statistical parameter $\alpha$, and
$d=1,2,3$. We explicitly note that the dilute limit $\lambda^d N/V \ll 1$
corresponds to $w_0 \gg 1$ ($\mu/\kB T \to -\infty$). We thus recover a monotonic
relation, as expected, which for $\alpha=0$ (Bose limit) leads to an
infinitesimally small negative chemical potential, in the limit of large
density.

\subsection{Pair correlation function}
\label{ssec:paircorr}

Following an analogous procedure, and performing the integration required in
Eq.~(\ref{eq:gnFourier1}) for an isotropic system in $d$ dimensions, for
the Fourier transform of the equilibrium distribution function of an assembly of
noninteracting exclusons we find
\begin{equation}
\lambda^d \tilde{n}(x) = x^{-\nu} \int_{w_0}^\infty dw \,
\frac{1}{w(1+w)} \bar{\epsilon}^{\nu/2} J_{\nu} \left( 2x
\bar{\epsilon}^{1/2} \right) ,
\label{eq:tildenx}
\end{equation}
where $x=\sqrt{\pi} r/\lambda$, $\nu=(d-2)/2$, and $J_\nu (z)$ is a Bessel
function of first kind, arising from the integration over angles in $d$
dimensions. Making use of the asymptotic properties of the Bessel functions
\cite{GR}, it can be shown that Eq.~(\ref{eq:tildenx}) correctly reduces to
Eq.~(\ref{eq:mud1}) for the particle density in the limit $x\to0$. Further
analytical results concerning the local excluson density $\tilde{n}(x)$ are derived
in App.~\ref{app:localdensity}.

\begin{figure}[t]
\centering
\includegraphics[height=0.9\columnwidth,angle=-90]{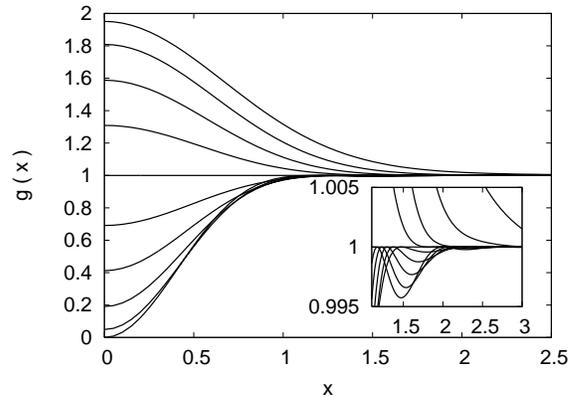}
\caption{Pair correlation function $g(x)$ (in scaled units),
Eq.~(\ref{eq:gnFourier}), in $d=2$, for $w_0 = 0.1$ (dense limit) and
$\alpha=0-1$ (top to bottom). Inset shows Friedel oscillations, which are
present only close to the FD limit ($0.5 < \alpha \leq 1$).}
\label{fig:pc1}
\end{figure}

Figure~\ref{fig:pc1} shows $g(x)$ in $d=2$ for $w_0 = 0.1$ (corresponding to a
relatively large particle density), and for $\alpha=0-1$. In the fermion limit
($\alpha=1$) and close to it ($0.5\lesssim\alpha\leq 1$), the pair correlation
function exhibits a correlation `hole' around $x=0$, whose depth decreases with
decreasing $\alpha$ and eventually vanishes as $\alpha\to\frac{1}{2}$. In the
same range of values of the statistical parameter, $g(x)$ is characterized by
damped `Friedel' oscillations (cf. inset of Fig.~\ref{fig:pc1}), which are more
pronounced close to the fermion limit. These oscillations are absent within
numerically accuracy in the prevalently bosonic range of the statistical
parameter, $0\leq \alpha \lesssim 0.5$, where the pair correlation function
displays a monotonically decreasing behavior, with correlations `piling up' at
$x=0$, in contrast with the fermion limit. Qualitatively similar results (not
shown here) are obtained in $d=1$ and $d=3$.

\begin{figure}[t]
\centering
\includegraphics[height=0.9\columnwidth,angle=-90]{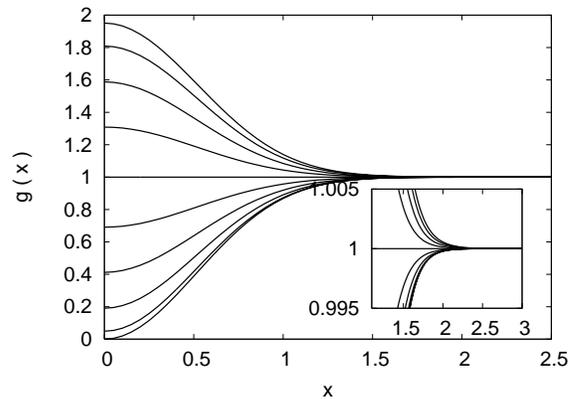}
\caption{Pair correlation function $g(x)$ (in scaled units),
Eq.~(\ref{eq:gnFourier}), in $d=2$, for $w_0 = 0.5$ (dilute limit) and
$\alpha=0-1$ (top to bottom). Inset shows the absence of Friedel oscillations,
even close to the FD limit ($0.5 \lesssim \alpha \leq 1$).}
\label{fig:pc2}
\end{figure}

Figure~\ref{fig:pc2} shows again the pair correlation function in $d=2$ for
$\alpha=0-1$, but now for $w_0 = 0.5$, corresponding to a lower particle
density. While the fermion side ($0.5\lesssim\alpha\leq 1$) is still
characterized by a correlation hole, and the boson side displays a monotonic
decrease of correlations, with a maximum at $x=0$, `Friedel' oscillations are
now beyond graphical resolution even in the quasi-fermion regime (cf. inset of
Fig.~\ref{fig:pc2}), as an effect of the reduced particle density. Again,
qualitatively similar results obtain also in $d=1$ and $d=3$.

\begin{figure}[t]
\centering
\includegraphics[height=0.9\columnwidth,angle=-90]{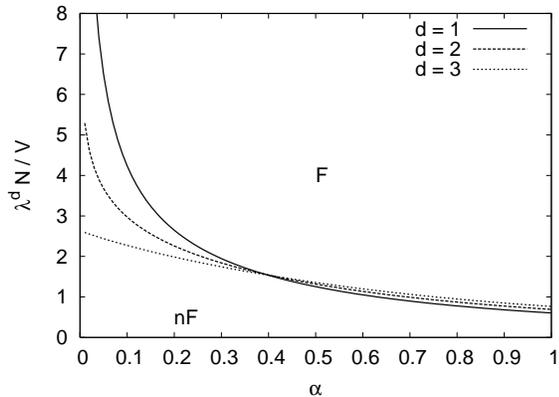}
\caption{Phase diagram in the plane of scaled density $\lambda^d N/V$ \emph{vs}
statistical parameter $\alpha$. The phase diagram features regions where
pronounced `Friedel' oscillations occur in the pair correlation function (F),
and where such oscillations are absent, or beyond graphical resolution (nF).
Different lines refer to various dimensionalities ($d=1,2,3$).}
\label{fig:friedel}
\end{figure}

In order to make more quantitative the latter statement, we now inquire on the
condition giving rise to `Friedel' oscillations in the pair correlation
function. Generally speaking, Friedel oscillations in many-body properties of
fermion assemblies arise from the presence of a discontinuity in the Fermi
distribution function at the Fermi level. Besides decreasing at $T=0$ as a
function of distance with a power law, depending on dimensionality, such
oscillations are then exponentially damped with a characteristic decay length
$\sim\lambda^2 \kF$, with $\kF$ the Fermi wavevector.

In the case of particles obeying fractional exclusion statistics, Wu's
distribution function Eq.~(\ref{eq:Wun}) in scaled units displays a smoother
inflection point at a value $\bar{\epsilon}_\ast$ corresponding to $w_\ast =
\alpha-1 + \sqrt{\alpha^2 - \alpha +1}$ in Eq.~(\ref{eq:walpha}). When $w_\ast
\geq w_0$, such an inflection point lies within the integration range in
Eq.~(\ref{eq:tildenx}), and does therefore give rise to marked `Friedel'
oscillations in the behavior of the pair correlation function. Such a condition
has been studied numerically as a function of scaled particle density $\lambda^d
N/V$ and statistical parameter $\alpha$, and is reproduced in
Fig.~\ref{fig:friedel} for various dimensionalities $d=1,2,3$ in the form of a
phase diagram.

The plane of scaled density $\lambda^d N/V$ \emph{vs} statistical parameter
$\alpha$ is divided into two regions by the line implicitly defined by the
equation $w_\ast = w_0$. At a fixed value of the statistical parameter $\alpha$,
say, one might think of increasing the fermionic-like correlations of a given
$g$-on assembly (\emph{i.e.} sharpening the `Friedel' oscillations in the pair
correlation function) by suitably increasing its scaled density $\lambda^d N/V$,
\emph{i.e.} increasing density $N/V$ or reducing temperature $T$.

Within this picture, one may also estimate the wavelength $\Lambda$ of the
oscillations in the pair correlation function with the approximate condition
$2\sqrt{\bar{\epsilon}_\ast} \Lambda \approx 2 \pi$, which straightforwardly leads to
\begin{equation}
\Lambda^2 \approx \frac{2\pi^2 \hbar^2}{m\epsilon_\ast} ,
\end{equation}
where all dimensions have been restored. Clearly, as $\epsilon_\ast \to 0$ as
$\alpha\to0$, one recovers the absence of `Friedel' oscillations in the bosonic
limit.

\section{Summary}
\label{sec:summary}

A brief discussion of the different concepts of fractional exchange and
fractional exclusion statistics is first given. Then we have focussed on some
thermodynamical properties and on the pair correlation function of an assembly
of non-interacting particles obeying exclusion statistics in arbitrary
dimensionality $d=1,2,3$ and given temperature $T$. In particular,
Eqs.~(\ref{eq:mud1}) and (\ref{eq:w0}) allow the chemical potential to be
studied as a function of scaled particle density and dimensionality. The pair
correlation function is then studied as a function of interparticle distance for
different values of the statistical parameter $\alpha$ and of the scaled density
$\lambda^d N/V$, where $\lambda$ is the de~Broglie thermal wavelength. The
behavior of the pair correlation function correctly exhibits a correlation
`hole' in the fermion limit, and a pronounced maximum in the boson limit. Away
from the boson limit, it also features `Friedel' oscillations, whose sharpness
and wavelength can be tuned as a function of the statistical parameter and of
scaled particle density. Our results show that, in principle, tuning density and
temperature may sharpen the statistical correlations among non-interacting
exclusons, thus ending in a behavior closer to either the fermionic or the
bosonic limit, depending on dimensionality and on the value of the statistical
parameter $\alpha$. This may be relevant to systems exhibiting the fractional
quantum Hall effect \cite{Haldane:91}, where anyons, \emph{i.e.} particles
obeying fractional exchange statistics, can be regarded as effectively
interacting exclusons \cite{Nayak:94}.

\begin{acknowledgments}
GGNA and NHM acknowledge that this paper was brought to completion during their
stay at the Centro di Ricerca Matematica ``Ennio De Giorgi'' of the Scuola
Normale Superiore, Pisa, Italy. Therefore they thank Professor M. P. Tosi and
the Director of the Centre, Professor M. Giaquinta, for much hospitality and for
the stimulating environment. 
\end{acknowledgments}

\appendix

\section{Recursion formulas for the local excluson density}
\label{app:localdensity}

Eq.~(\ref{eq:tildenx}) yields $\tilde{n}_\nu (x,w_0 ) \equiv \tilde{n} (x)$ as a
function of scaled interparticle distance $x$ and generalized inverse fugacity
$w_0$, at a given statistical parameter $\alpha$ and reduced dimensionality
$\nu=(d-2)/2$. By differentiating Eq.~(\ref{eq:tildenx}) with respect to $x$,
and making use of the appropriate recursion formula for the Bessel functions
\cite{GR}, one obtains
\begin{equation}
\frac{\partial \tilde{n}_\nu}{\partial x} + 2 x \tilde{n}_{\nu+1} = 0.
\label{eq:pdex}
\end{equation}
Similarly, by differentiating the same Eq.~(\ref{eq:tildenx}) with respect to
$w_0$, and taking into account Eq.~(\ref{eq:walpha}), one obtains
\begin{equation}
\frac{\partial \tilde{n}_\nu}{\partial w_0} + \left( \frac{d\bar{\epsilon}}{dw}
\right)_0 \tilde{n}_{\nu-1} + \rho_\nu (w_0 ) = 0,
\label{eq:pdew0}
\end{equation}
where
\begin{equation}
\left( \frac{d\bar{\epsilon}}{dw} \right)_0 = \frac{\alpha + w_0}{w_0 (1+w_0 )}
,
\end{equation}
and
\begin{equation}
\rho_\nu (w) = \frac{1}{\Gamma(\nu+1)} \frac{1}{w(1+w)} \bar{\epsilon}^\nu
\end{equation}
is the density of states in the $w$ variable, in $d=2\nu+2$ dimensions.
Eqs.~(\ref{eq:pdex}) and (\ref{eq:pdew0}) are then differential recursion
formulas relating the local excluson density at different dimensionalities.
Combining Eqs.~(\ref{eq:pdex}) and (\ref{eq:pdew0}), and noting that
$\rho_{\nu+1} (w_0 ) = \lim_{w\to w_0} \rho_{\nu+1} (w) = 0$ for $d=1,2,3$, one
obtains
\begin{equation}
\frac{\partial^2 \tilde{n}_\nu}{\partial w_0 \partial x} = 2x \frac{\alpha +
w_0}{w_0 (1+w_0 )} \tilde{n}_\nu ,
\end{equation}
which is a hyperbolic partial differential equation for the local excluson
density, similar to Klein-Gordon equation, but with a variable `mass' term.

\begin{small}
\bibliographystyle{apsrev}
\bibliography{a,b,c,d,e,f,g,h,i,j,k,l,m,n,o,p,q,r,s,t,u,v,w,x,y,z,zzproceedings,Angilella}
\end{small}

\end{document}